\documentclass[epj]{svjour}
\usepackage{graphics}
\begin{document}
\title{A Machian Solution of the Hierarchy Problem}
\author{Merab Gogberashvili}
\institute{Andronikashvili Institute of Physics, 6 Tamarashvili
St., Tbilisi 0177, Georgia \and Javakhishvili State University, 3
Chavchavadze Avenue, Tbilisi 0128, Georgia}
\date{Received:   }
\abstract{The new interpretation of Mach's principle of mass of a
particle being a measure of the interactions of this particle with
all other gravitating particles inside its causal spheres is
introduced. It is shown that within some alternative model of
gravitation that incorporates this principle, the Machian
influence of the universe can reduce Planck's scale to the
electro-weak scale and the large number that is needed to explain
the hierarchy between the scales is the amount of gravitating
particles inside the universe horizon. Our model can lead to new
observable effects at cosmological distances and close to the
sources of a strong gravitational field.
 }
\maketitle

%%%%%%%%%%%%%%%%%%%%%%%%%%%%%%%%%%%%%%%%%%%%%%%%%%%%%%%%%%%%%%%%%%

In particle physics the hierarchy problem mainly consists in the
existence of a large separation between the electro-weak and
Planck's scales, or equivalently between the Higgs and Planck
masses. The Higgs mass is quadratically sensitive to the cut-off
of the theory and if the cut-off is very high it is not clear what
stabilizes it. If the cut-off is not far from the electro-weak
scale, then one has to explain why gravity is so weak. In both
cases, the solution should come up in the form of the large number
$N$ that provides the hierarchy between the Planck mass $M_P$ and
the Higgs mass $M_H$,
\begin{equation} \label{P=NH}
M_P^2 \approx  N  M_H^2~.
\end{equation}
For example, in SUSY the large number $N$ is the ratio of Planck's
mass to the supersymmetry breaking scale, whereas in large extra
dimensions scenarios the large volume of the extra space sets the
hierarchy.

In the recent papers \cite{DGKN,Dvali} there was considered a
different approach, in which $N$ in (\ref{P=NH}) is the number of
species of some new quantum fields. The argument of \cite{DGKN} is
that each $N$ species of fields, with masses at the scale $M_H$,
coupled to gravity will contribute to the renormalization of
Planck's mass the factor $\sim M_H^2$. Neglecting the accidental
cancellations this has to be multiplied by the number of species.
As a result the effective contribution to the Planck mass is $\sim
N M_H^2$. Later in \cite{Dvali} also a non-perturbative argument
was introduced: in the model with a large amount of different
quantum fields Planck's mass violating the bound (\ref{P=NH}) is
inconsistent with black hole physics.

Here we want to elaborate another idea of how to reduce Planck's
scale to the electro-weak scale. The model based on Mach's effect
asserting that any local gravitation interaction is affected by
all matter in the universe \cite{Mach}. We shall show that Mach's
effect can reduce the local strength of the gravitational
interaction, and the large number $N$ in (\ref{P=NH}) can be the
amount of gravitating particles within the universe horizon. Note
that up to astrophysical distances this model does not requires
radical changes of standard physics, such as the introduction of
extra dimensions, or new interactions.

Mach's principle was considered by Einstein in the development of
general relativity \cite{Ein} but was rejected when he found that
his field equations admit curved vacuum and asymptotically flat
solutions. Mach's principle deals with a fundamental issue of
physics: the origin of inertia (or the nature of inertial mass).
General relativity does not explain the origins of inertial and
gravitational masses; it just states that they are equivalent.
According to Mach some close mass, like a galaxy, should lead to
anisotropy of the inertial mass $m_i$ of a body. However, precise
tests do not show this difference; for example, it was found that
$\Delta m_i/m_i \leq 10^{-20}$ for nucleons \cite{delta-m}. So the
existing data are in favor of Einstein's equivalence principle and
against a classical interpretation of Mach's effect. Discussions
on the present status of Mach's principle can be found in
\cite{Mach+}.

In our opinion contradictions with Einstein's equivalence
principle can be resolved if we use Mach's ideas on the level of
particle physics assuming strong non-locality in a future quantum
theory of gravity. We suppose that the mass (inertial and
gravitational) of any particle is a measure of the interactions of
this particle with all other gravitating particles inside its
causal sphere and not with some classical massive objects like
distant stars. Since for the distances of the order of the horizon
the distribution of matter is isotropic (cosmological principle),
local fluctuations of the gravitational potential (or the
existence of close masses) does not affect the description of the
inertial mass of a particle.

In particle physics, mass is a measure of the coupling of
particles with the Higgs field and it is usually assumed that all
particles in the universe receive there masses by the same
mechanism. It is natural to suppose that both the inertial and
gravitational masses are connected with the fundamental Higgs
scale, and Newton's constant is some emergent quantity. So in the
model that incorporates Mach's principle in a novel interpretation
the universal background Higgs field probably can be replaced by
the background gravitational potential of the universe.

The main observation is that the inertial $m_i$ and the
gravitational $m_g$ masses in general are not necessarily
equivalent. One can state only that these masses of an object are
proportional to one another:
\begin{equation} \label{m=Nm}
m_g = \sqrt{N} m_i~,
\end{equation}
where $N$ is some constant of proportionality. This is usually
referred as the 'weak form' of the equivalence principle. The
assumption (\ref{m=Nm}) does not means a violation of the
equivalence principle since $N$ is still the same for all bodies,
whatever their compositions are. In Einstein's theory it is
assumed that inertial and gravitational masses of any object are
exactly equivalent, i.e. $N=1$. However, all the results of
standard physics would still remain valid if we write (\ref{m=Nm})
and at the same time introduce the new gravitational constant
\begin{equation} \label{g=NG}
g =  N G~,
\end{equation}
where $G$ is usual Newton's constant.

Equation (\ref{g=NG}) is a relation similar to (\ref{P=NH}). We
assume that, because of Machian effects, the constant $g$ in
(\ref{g=NG}) can be connected with the electro-weak scale, i. e.
\begin{equation}
g \sim \frac 1{M_H^2}~,
\end{equation}
and the large quantity $N$ is the number of gravitating particles
within the horizon that are sources of gravitational interaction.
If this mechanism works it can explain some large number
coincidence in physics. A recent review on the large numbers
hypothesis, proving the existence of some deep connection between
micro and macro physics, the reader can find in \cite{large}.

Another observation is that mass is a positive quantity and
screening of gravity is impossible. The 'gravitational flux' of a
body should be conserved and this flux will distribute to all
interacting particles. So in any two-body process the influence of
all other matter inside their causal sphere should be taken into
account. Because of the existence of a large amount of matter in
the universe this should effectively reduce the strength of any
two-body interaction, i.e. the gravitational constant.

When considering the gravitational interaction of close objects
one can replace the distant universe by a spherical shell of the
effective mass $M$ and the effective radius $R$. This physical
substitution is analogous to replacing of a spherical mass by a
point mass. As in the electrical case this shell acts similar to a
gravitational Faraday cage inside of which a huge, yet constant,
gravitational potential exists,
\begin{equation} \label{phi-U}
\phi = - \frac{M G}{R}~.
\end{equation}
Since the potential (\ref{phi-U}) is constant the 'universe field'
$E$ on an inertial particle is zero,
\begin{equation}
\nabla \phi = 0~.
\end{equation}
However, accelerated particles feel the constant 'universe
potential' (\ref{phi-U}) and similar to the induction law in
electrodynamics for the field strength we have \cite{Sia}
\begin{equation} \label{E-U}
E =  - \nabla \phi - \frac{\phi}{c^2}~ a  = - \frac{\phi}{c^2}
~a~,
\end{equation}
where $a$ is the vector of acceleration and $c$ is the speed of
light.

In the case of a homogeneous and isotropic matter distribution in
the universe with the average density $\rho$ we have the following
relationships:
\begin{equation}\label{phi=c}
\frac{\phi}{c^2} = - \frac{G\rho V}{Rc^2} = - \frac{2 \pi G \rho
R^2}{c^2} = -\frac{2 \pi G \rho}{H^2} ~.
\end{equation}
For simplicity the considered volume of the universe is a Hubble
sphere of radius
\begin{equation}
R = \frac cH~.
\end{equation}
Using the formula for the critical mass density of the universe
\begin{equation} \label{rho-c}
\rho_{c} = \frac{3H^2}{8\pi G}
\end{equation}
from (\ref{phi=c}) we found that the relationship
\begin{equation} \label{equality}
\phi \approx  - c^2
\end{equation}
is valid to a reasonable degree of precision for our simple
considerations. It seems that the relation (\ref{equality}) is
valid for all stages of the universe expansion and the
gravitational potential in the universe remains unchanged
(conserved) since the Planck time \cite{Ken},
\begin{equation}
\phi = - \frac{G M_P}{ l_P} = - c^2~,
\end{equation}
where $l_P$ is Planck's length. This formula in fact coincides
with the definition of Newton's constant by Planck's mass
\begin{equation}
G \sim \frac 1{M_P^2}
\end{equation}
in units where $c = 1$.

The gravitational field of the universe (\ref{E-U}) on a particle
with the mass $m_g$ due to (\ref{equality}) results in the
standard expression of the inertial force,
\begin{equation} \label{F}
F = m_g E = m_i a~,
\end{equation}
where the notion of inertial mass, as the measure of the
gravitational interaction with the universe, was introduced by
\begin{equation} \label{mi=mg}
m_i = - m_g \frac{\phi}{c^2}~.
\end{equation}
This formula provides support to Mach's hypothesis about the
origin of inertia.

One can also obtain (\ref{equality}) from Friedman's equations by
the integration of the deceleration vector along the radius of the
causal sphere from the center to the particle horizon \cite{Sig}.
The relation (\ref{equality}) can be understood as the definition
of the horizon of the universe - the internal horizon of the
Schwarzschild sphere, or alternatively as the distance at which
the velocity of recession of galaxies is equal to the speed of
light.

The relation (\ref{equality}) leads also to another conclusion
that the total energy (inertial plus gravitational) of a particle
at rest to the universe is zero \cite{Sia},
\begin{equation} \label{Ei=Eg}
m_ic^2 + m_g \phi = 0~.
\end{equation}
One can show that this relation is equivalent to the standard
definition of inertial coordinate system as the frame in which all
forces on a body compensate.

The fact of conservation of the potential $\phi$ during expansion
raises the question of the variation of the gravitational constant
$G$, since only in a very specific cosmological model the mass of
the universe $M$ increases linearly with the radius $R$. While the
macroscopic characteristics of the universe can vary during the
expansion, it is usually assumed that the parameters of particle
physics are unaffected. So if $g$ in (\ref{g=NG}) is connected
with the electro-weak scale it should remain constant, but the
number $N$ and Newton's constant $G$ can change in time. The
possibility of a variation of $G$ was considered by many authors
(for a review see \cite{Uzan}).

To estimate $N$, and thus $g$, let us consider the spherical model
universe of radius $R$ consisting of $N$ uniformly distributed
particles of gravitational mass $m_g$. Each particle in our toy
model gravitationally 'feels' all the other $(N-1)$ particles. The
mean contribution of each pair to the gravitational energy will be
$m_g^2/R$. Thus the 'universe potential' will contain $\approx
N^2$ terms formed by all pairs of particles, for which the mean
separation will be $R$, i. e.
\begin{equation} \label{phi-N2}
\phi \approx - \frac{m_g G}{R}~N^2~.
\end{equation}
Comparing this formula with (\ref{phi-U}), we see that due to the
interactions the active gravitational mass of the universe in this
model is equal to
\begin{equation} \label{M}
M = N^2 m_g~,
\end{equation}
and not to
\begin{equation}
\int \rho dV = N m_g
\end{equation}
as expected from Gauss's law in the additive case.

From the relation (\ref{phi-N2}) we see that for small values of
$N$, if we state that $m_i = m_g$ and $M = Nm_g$, gravity gets
stronger by the factor $N$. Thus the gravity scale increases with
decreasing of $N$ according to the law (\ref{g=NG}) and achieves
its maximal value $g$ in the case of the presence of only two
particles in the universe. This means that the real interaction
scale in this model is $g \sim 1/M_H^2$ and $G$ is some emergent
quantity incorporating the influence of all particles on any local
interaction.

To estimate the number $N$ we need to calculate the amount of
gravitating particles (including dark components) within the
Hubble horizon. The inertial mass of a typical particle, for
example a proton, has the order $m_p \sim 10^{-27} kg$. From the
estimations of the critical mass density (\ref{rho-c}), it is
known that $M \sim 10^{53} kg$. From (\ref{M}) it follows that the
gravitational mass of the universe $M$ is $N$ times larger than
the sum of the masses of all particles. So the proton equivalent
of the total number of gravitating particles is
\begin{equation}
N \sim \sqrt{\frac{M}{m_p}} \sim 10^{40}~.
\end{equation}
This is a large number of the order that one needs to reduce
Planck's scale to the electro-weak scale by the formula
(\ref{g=NG}).

Another toy model that shows that $g$ in (\ref{g=NG}) really can
be connected with the electro-weak scale is the universe filled
with $N$ particles of the masses $m_g$ half of which carry
positive unit charges $e$ and the other half negative charges $-e$
($N$ is an even integer) \cite{Bar}. It is known that the
intensity of electromagnetic interaction of two elementary
particles is $10^{40}$ times larger than their gravitational
interaction for the same distance,
\begin{equation}
\frac{ke^2}{Gm_g^2} \sim 10^{40}~,
\end{equation}
where $k$ is Coulomb's constant. However, to compare the total
intensities one should take into account the gravitational
interaction of all particles in the universe. In estimations of
the potentials the fact that all masses are positive, while half
the charges are positive and half negative leads to large
differences. The gravitational potential expressed by
(\ref{phi-N2}) contains $\approx N^2$ terms. In contrast, each
charge $e$ beyond the Debye radius finds itself in a neutral
cloud. In fact, the total charge outside any considered charge $e$
will be $-e$, and the 'mean distance' of the opposite charge will
be $R$. Therefore, the electrostatic potential consists of $N$
terms, each of magnitude $\approx e^2/R$. So the fundamental
gravitational interaction engages all $N$ particles and only two
electromagnetic ones. Thus the ratio of the total gravitational
and electromagnetic intensities is close to unity, i. e.
\begin{equation}
\frac{ke^2}{gm_g^2} \sim 1~.
\end{equation}
From this estimation we can conclude that in this model the
Coulomb constant $k$ and the fundamental gravitational constant
$g$ are connected to the same electro-weak scale.

\medskip

{\bf Discussions:} In this paper the new interpretation of Mach's
principle that the mass of a particle is a measure of the
interactions of this particle with all other gravitating particles
inside its causal sphere was introduced. It was shown that within
some alternative theory of gravitation, which assumes this
non-local Machian response of the universe in local experiments,
one can explain the observed hierarchy between the scales of
particle physics and gravity. The reason why gravity seems to be
weak in two-body interactions is that the mass (the charge of
gravity) is always positive. Then screening of gravity is
impossible and the gravitational flux of any body is distributed
over all matter inside the horizon. It was found that in this
model the large number needed for effective weakening of the
gravitational constant from the electro-weak to Planck's scale
coincides with the amount of gravitating particles in the
universe. Note that, unlike other models with a single scale for
particle physics and gravity, the incorporation of Mach's
principle does not requires radical changes of standard physics up
to astrophysical distances.

Since we want to reduce the fundamental scale of gravity, the
model presented here can lead to observable new effects in high
energy physics experiments, or for strong gravitational fields. In
the cosmological case the modification of the description of mass
can imitate the effects of dark energy at ultra-large distances.

%%%%%%%%%%%%%%%%%%%%%%%%%%%%%%%%%%%%%%%%%%%%%%%%%%%

\begin{acknowledgement}
I would like to thank Prof. Z. Berezhiani for helpful discussions.
\end{acknowledgement}

%%%%%%%%%%%%%%%%%%%%%%%%%%%%%%%%%%%%%%%%%%%%%%%%%%%%

\end{document}